\documentclass[12pt]{article}
 \usepackage{epsfig}
 \def\be{\begin{equation}}
 \def\ee{\end{equation}}
 \def\bea{\begin{eqnarray}}
 \def\eea{\end{eqnarray}}
 \usepackage{graphicx}% Include figure files

 \catcode`\@=11
 \def\lsim{\mathrel{\mathpalette\@versim<}}
 \def\gsim{\mathrel{\mathpalette\@versim>}}
 \def\@versim#1#2{\vcenter{\offinterlineskip
 \ialign{$\m@th#1\hfil##\hfil$\crcr#2\crcr\sim\crcr } }}
 \catcode`\@=12

 \catcode`@=12
\begin{document}
 \thispagestyle{empty}
 \begin{flushright}
 UCRHEP-T623\\
 May 2022\
 \end{flushright}
 \vspace{0.6in}
 \begin{center}
 {\LARGE \bf Type III Neutrino Seesaw,\\ 
Freeze-In Long-Lived Dark Matter,\\
and the $W$ Mass Shift\\}
 \vspace{1.5in}
 {\bf Ernest Ma\\}
 \vspace{0.1in}
{\sl Department of Physics and Astronomy,\\ 
University of California, Riverside, California 92521, USA\\}
\end{center}
 \vspace{1.2in}

\begin{abstract}\
In the framework of seesaw neutrino masses from heavy fermion triplets 
$(\Sigma^+,\Sigma^0,\Sigma^-)$, the addition of a light fermion singlet $N$ 
and a heavy scalar triplet $(\rho^+,\rho^0,\rho^-)$ has some important 
consequences.  The new particles are assumed to be odd under a new $Z_2$ 
symmetry which is only broken softly, both explicitly and spontaneously. 
With $N-\Sigma^0$ mixing, freeze-in long-lived dark matter through Higgs 
decay becomes possible.  At the same time, the $W$ mass is shifted slightly 
upward, as suggested by a recent precision measurement.
\end{abstract}

\newpage
\baselineskip 24pt

\noindent \underline{\it Introduction}~:~ 
The well-known seesaw mechanism for tiny Majorana neutrino masses has three 
simple tree-level realizations~\cite{m98}, depending on the heavy 
intermediary particles involved.  Whereas Type I [heavy fermion singlet $N$] 
and Type II [heavy scalar triplet $(\xi^{++},\xi^+,\xi^0)$] are routinely  
considered in numerous papers, Type III [heavy fermion triplet 
$(\Sigma^+,\Sigma^0,\Sigma^-)$] is rather less studied~\cite{flhj89,m09}.
In this paper, in addition to three copies of heavy $\Sigma$ for obtaining 
three light Majorana neutrinos, a new $Z_2$ symmetry is assumed under which 
one new \underline {\it light} singlet fermion $N$ and one new heavy scalar 
triplet $(\rho^+,\rho^0,\rho^-)$ are odd, and all other fields are even.
Whereas all dimension-four terms of the resulting Lagrangian must respect 
$Z_2$ [which forbids $N$ from coupling to lepton doublets through the usual 
Higgs doublet $\Phi$ of the standard model of quarks and leptons (SM)], this 
symmetry is broken explicitly by the dimension-three soft term 
$\Phi^\dagger \rho \Phi$, resulting thus in a small nonzero 
vacuum expectation value $v_\rho$ for $\rho^0$.  

Three important consequences follow.  (A) The $W$ mass gets a slight upward 
shift~\cite{cdf22}.  (B) The $\phi^0-\rho^0$ and $\Sigma^0-N$ mixings allow 
the SM Higgs boson $h$ to decay to $NN$.  (C) $N$ decays through its 
mixing with the heavy $\Sigma^0$ which couples to $\nu \phi^0$, converting 
thereby to $\nu \bar{f} f$, where $f$ is the heaviest fermion kinematically 
allowed.  Hence $N$ is possibly a long-lived dark-matter candidate, produced 
in a freeze-in scenario~\cite{hjmw10} through rare $h$ decay~\cite{m19}. 

\noindent \underline{\it Higgs Potential}~:~
The Higgs potential $V$ of this proposal consists of the SM Higgs doublet 
$\Phi$ and the new real scalar triplet $\rho$ which is odd under the 
assumed $Z_2$, i.e.
\begin{eqnarray}
V &=& -\mu_0^2 (\Phi^\dagger \Phi) + {1 \over 2} m_1^2 (\vec{\rho} \cdot 
\vec{\rho}) + {1 \over 2} \lambda_1 (\Phi^\dagger \Phi)^2 + {1 \over 8} 
\lambda_2 (\vec{\rho} \cdot \vec{\rho})^2 \nonumber \\ 
&+& {1 \over 2} \lambda_3 (\Phi^\dagger \Phi)(\vec{\rho} \cdot \vec{\rho}) + 
\sqrt{2} \mu_1 \Phi^\dagger (\vec{\sigma} \cdot \vec{\rho}) \Phi,
\end{eqnarray}
where the $\mu_1$ trilinear term breaks $Z_2$ softly.  Let
\begin{equation}
\phi^0 = {1 \over \sqrt{2}} (v_0 + h), ~~~ \rho^0 = v_1 + s,
\end{equation}
then $v_{0,1}$ are determined by
\begin{eqnarray}
0 &=& -\mu_0^2 + {1 \over 2} \lambda_1 v_0^2 + {1 \over 2} \lambda_3 v_1^2 
- {\mu_1 v_1 \over \sqrt{2}}, \\ 
0 &=& v_1 \left( m_1^2 + {1 \over 2} \lambda_2 v_1^2 + {1 \over 2} 
\lambda_3 v_0^2 \right) - {\mu_1 v_0^2 \over 2\sqrt{2}}.
\end{eqnarray}
For large and positive $m_1^2$, the scalar seesaw solution~\cite{m01} is 
\begin{equation}
v_0^2 \simeq {2 \mu_0^2 \over \lambda_1}, ~~~ v_1 \simeq {\mu_1 v_0^2 \over 
2\sqrt{2} m_1^2}.
\end{equation}
The $2 \times 2$ mass-squared matrix spanning $h$ and $s$ is then
\begin{equation}
{\cal M}^2_{hs} \simeq \pmatrix{ \lambda_1 v_0^2 & -\mu_1 v_0/\sqrt{2} 
\cr -\mu_1 v_0/\sqrt{2} & m_1^2},
\end{equation}
with $h-s$ mixing given by
\begin{equation}
\theta_{hs} \simeq {\mu_1 v_0 \over \sqrt{2}m_1^2} \simeq {2 v_1 \over v_0}.
\end{equation}
To explain the new precision measurement of the $W$ mass~\cite{cdf22}, i.e. 
\begin{equation}
M_W = 80.4335 \pm 0.0094~{\rm GeV},
\end{equation}
which is several standard deviations above the prediction of the SM 
($v_1=0$), a central value of $v_1 \simeq 3.68$ GeV may be extracted 
from the analysis of Ref.~\cite{ps22}.

\noindent \underline{\it Singlet-Triplet Fermion Mixing}~:~
Neutrinos obtain seesaw masses through the heavy fermion triplets from 
the Yukawa couplngs
\begin{equation}
{\cal L}_Y = \sqrt{2} f_\nu (\bar{\nu},\bar{l})_L (\vec{\sigma} \cdot 
\vec{\Sigma}_R) \pmatrix{\bar{\phi}^0 \cr -\phi^-},
\end{equation}
resulting in the Dirac mass $m_{\nu\Sigma} = f_\nu v_0/2$, and then the usual 
seesaw Majorana neutrino mass $m_\nu = f_\nu^2 v_0^2/4m_\Sigma$.  The $\nu-S$ 
mixing is $\sqrt{m_\nu/m_\Sigma}$, and the coupling of $\Sigma$ to 
$\nu h$ is $\sqrt{m_\nu m_\Sigma}/v_0$.

Since $N$ is odd under $Z_2$, it does not couple to $\nu$ through $\phi^0$. 
However, it does couple to $\Sigma$ through $\rho$, i.e.
\begin{equation}
{\cal L}'_Y = f_N \bar{N}_L (\vec{\rho} \cdot \vec{\Sigma}_R).
\end{equation}
The $N-\Sigma^0$ mixing is then $f_N v_1/m_\Sigma$ and the $s$ coupling 
to $NN$ is $f_N^2 v_1/m_\Sigma$.

\noindent \underline{\it Higgs Decay to $NN$}~:~
Since $N$ is a singlet fermion, the Higgs boson $h$ does not couple to $NN$ 
directly.  It does so first through $h-s$ mixing, then through $N-\Sigma^0$ 
mixing, as shown in Fig.~1.  
\begin{figure}[htb]
 \vspace*{-6cm}
 \hspace*{-3cm}
 \includegraphics[scale=1.0]{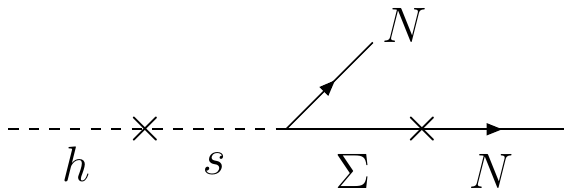}
 \vspace*{-22.0cm}
 \caption{Decay of $h$ to $NN$.}
 \end{figure}

\noindent The effective coupling is then
\begin{equation}
f_h \simeq \left( {2 v_1 \over v_0} \right) \left( {f_N^2 v_1 \over m_\Sigma} 
\right) = {2 f_N^2 v_1^2 \over v_0 m_\Sigma}.
\end{equation}
The decay rate of $h \to NN + \bar{N}\bar{N}$ is~\cite{mr21}
\begin{equation}
\Gamma_h = {f_h^2 m_h \over 8 \pi} \sqrt{1-4r^2}(1-2r^2),
\end{equation}
where $r=m_N/m_h$.  Now $N$ is assumed light and a candidate for long-lived 
dark matter.  The correct relic abundance is obtained~\cite{ac13} if 
$f_h \sim 10^{-12} r^{-1/2}$, provided that the reheat temperature of the 
Universe is above $m_h$ but well below $m_\rho$ and $m_\Sigma$.

\noindent \underline{\it Long-Lived Dark Matter}~:~
The singlet fermion $N$ is assumed light and decays only through its 
mixing with $\Sigma^0$ which couples to $\nu h$.  Through the virtual 
Higgs, its coupling to $\nu \bar{f} f$ is then given by
\begin{equation}
G_N = {f_N v_1 \over m_\Sigma} \left( {\sqrt{m_\nu m_\Sigma} \over v_0} 
\right) {1 \over m^2_h} \left( {m_f \over v_0} \right),
\end{equation}
where $f$ is a fermion allowed kinematically in the decay, as shown 
in Fig.~2. The decay rate is analogous to that of muon decay, i.e.
\begin{equation}
\Gamma_N = {G_N^2 m_N^5 \over 48 (4\pi)^3}.
\end{equation}  
\begin{figure}[htb]
 \vspace*{-6cm}
 \hspace*{-3cm}
 \includegraphics[scale=1.0]{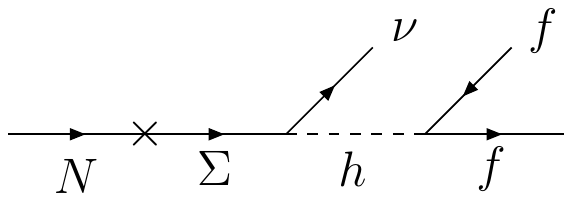}
 \vspace*{-22.0cm}
 \caption{Decay of $N$ to $\nu \bar{f} f$.}
 \end{figure}
\noindent  Consider for example $m_N = 0.1$ GeV, then 
$f_h \sim 10^{-12} (0.1/125)^{-1/2}$ from freeze-in Higgs decay implies
\begin{equation}
m_\Sigma/f_N^2 \sim 3.1 \times 10^9~{\rm GeV}
\end{equation}
for $v_1=3.68$ GeV.  In $N \to \nu \bar{f} f$ decay, only an electron-positron 
pair is possible, hence $m_f=m_e$ and
\begin{equation} 
G_N \sim 3.5 \times 10^{-22}~{\rm GeV}^{-2}
\end{equation}
for $m_\nu = 0.1$ eV. The $N$ lifetime is then 
\begin{equation}
\tau_N \sim 5.1 \times 10^{28}~{\rm s},
\end{equation}
many orders of magnitude greater than the age of the Universe and 
satisfies bounds from all cosmological considerations~\cite{sw17}.

\noindent \underline{\it Conclusion}~:~
In this paper, a first example of long-lived freeze-in dark matter is 
presented in the context of Type III seesaw neutrino masses using heavy 
fermion triplets $\Sigma$.  The key is the addition of a light fermion 
singlet $N$ and a real scalar triplet $\rho$, both odd under a softly 
broken $Z_2$ symmetry.  The rare decay of the SM Higgs to $NN$ accounts 
for the dark matter relic abundance of the Universe, with $N$ having a 
lifetime many orders of magnitude greater than the age of the Universe.  
This is accomplished with $m_N = 0.1$ GeV, $m_\Sigma \sim 10^9$ GeV, and 
$\langle \rho^0 \rangle = 3.68$ GeV, which also explains the shift in 
the $W$ boson mass, observed recently.

\noindent \underline{\it Acknowledgement}~:~
This work was supported in part by the U.~S.~Department of Energy Grant 
No. DE-SC0008541.  

\bibliographystyle{unsrt}

\end{document}